\documentstyle[12pt,epsfig,wrapfig]{article}
\catcode`\@=11
\long\def\@makefntext#1{
\protect\noindent \hbox to 3.2pt {\hskip-.9pt  
$^{{\ninerm\@thefnmark}}$\hfil}#1\hfill}                

\def\@makefnmark{\hbox to 0pt{$^{\@thefnmark}$\hss}}  
        
\def\ps@myheadings{\let\@mkboth\@gobbletwo
\def\@oddhead{\hbox{}
\rightmark\hfil\ninerm\thepage}   
\def\@oddfoot{}\def\@evenhead{\ninerm\thepage\hfil
\leftmark\hbox{}}\def\@evenfoot{}
\def\sectionmark##1{}\def\subsectionmark##1{}}

\renewcommand{\thefootnote}{\fnsymbol{footnote}}

\newcounter{sectionc}\newcounter{subsectionc}\newcounter{subsubsectionc}
\renewcommand{\section}[1] {\vspace*{0.6cm}\addtocounter{sectionc}{1} 
\setcounter{subsectionc}{0}\setcounter{subsubsectionc}{0}\noindent 
        {\normalsize\bf\thesectionc. #1}\par\vspace*{0.4cm}}
\renewcommand{\subsection}[1] {\vspace*{0.6cm}\addtocounter{subsectionc}{1} 
        \setcounter{subsubsectionc}{0}\noindent 
        {\normalsize\it\thesectionc.\thesubsectionc. #1}\par\vspace*{0.4cm}}
\renewcommand{\subsubsection}[1]
{\vspace*{0.6cm}\addtocounter{subsubsectionc}{1}
        \noindent {\normalsize\rm\thesectionc.\thesubsectionc.\thesubsubsectionc. 
        #1}\par\vspace*{0.4cm}}

\newcounter{appendixc}
\newcounter{subappendixc}[appendixc]
\newcounter{subsubappendixc}[subappendixc]

\renewcommand{\appendix}[1] {\vspace*{0.6cm}
        \refstepcounter{appendixc}
        \setcounter{figure}{0}
        \setcounter{table}{0}
        \setcounter{equation}{0}
        \renewcommand{\thefigure}{\Alph{appendixc}.\arabic{figure}}
        \renewcommand{\thetable}{\Alph{appendixc}.\arabic{table}}
        \renewcommand{\theappendixc}{\Alph{appendixc}}
        \renewcommand{\theequation}{\Alph{appendixc}.\arabic{equation}}
        \noindent{\bf Appendix \theappendixc #1}\par\vspace*{0.4cm}}

\def\abstracts#1{{
        \centering{\begin{minipage}{12.2truecm}\footnotesize\baselineskip=12pt\noindent
        \centerline{\footnotesize ABSTRACT}\vspace*{0.3cm}
        \parindent=0pt #1
        \end{minipage}}\par}} 

\renewenvironment{thebibliography}[1]
        {\begin{list}{\arabic{enumi}.}
        {\usecounter{enumi}\setlength{\parsep}{0pt}
\setlength{\leftmargin 1.25cm}{\rightmargin 0pt}
         \setlength{\itemsep}{0pt} \settowidth
        {\labelwidth}{#1.}\sloppy}}{\end{list}}

\newcounter{itemlistc}
\newcounter{romanlistc}
\newcounter{alphlistc}
\newcounter{arabiclistc}

\newcommand{\fcaption}[1]{
        \refstepcounter{figure}
        \setbox\@tempboxa = \hbox{\footnotesize Fig.~\thefigure. #1}
        \ifdim \wd\@tempboxa > 6in
           {\begin{center}
        \parbox{6in}{\footnotesize\baselineskip=12pt Fig.~\thefigure. #1}
            \end{center}}
        \else
             {\begin{center}
             {\footnotesize Fig.~\thefigure. #1}
              \end{center}}
        \fi}

\newcommand{\tcaption}[1]{
        \refstepcounter{table}
        \setbox\@tempboxa = \hbox{\footnotesize Table~\thetable. #1}
        \ifdim \wd\@tempboxa > 6in
           {\begin{center}
        \parbox{6in}{\footnotesize\baselineskip=12pt Table~\thetable. #1}
            \end{center}}
        \else
             {\begin{center}
             {\footnotesize Table~\thetable. #1}
              \end{center}}
        \fi}

\def\@citex[#1]#2{\if@filesw\immediate\write\@auxout
        {\string\citation{#2}}\fi
\def\@citea{}\@cite{\@for\@citeb:=#2\do
        {\@citea\def\@citea{,}\@ifundefined
        {b@\@citeb}{{\bf ?}\@warning
        {Citation `\@citeb' on page \thepage \space undefined}}
        {\csname b@\@citeb\endcsname}}}{#1}}

\newif\if@cghi
\def\cite{\@cghitrue\@ifnextchar [{\@tempswatrue
        \@citex}{\@tempswafalse\@citex[]}}
\def\citelow{\@cghifalse\@ifnextchar [{\@tempswatrue
        \@citex}{\@tempswafalse\@citex[]}}
\def\@cite#1#2{{$\null^{#1}$\if@tempswa\typeout
        {IJCGA warning: optional citation argument 
        ignored: `#2'} \fi}}

 1
 1
 1

\font\ninerm=cmr9

\begin{document}

\centerline{\normalsize\bf OBSERVATION OF LOW {\boldmath $x$} PHENOMENA IN HADRONIC 
FINAL STATES\footnote{To
appear in Proc.\ of ``New Trends in HERA Physics'', Ringberg, 
Tegernsee, Germany, May 1997.}}
\baselineskip=22pt
\centerline{\footnotesize 
G\"UNTER  GRINDHAMMER}
\baselineskip=13pt
\centerline{\footnotesize\it Max-Planck-Institut f\"ur Physik, 
(Werner-Heisenberg-Institut)}
\baselineskip=12pt
\centerline{\footnotesize\it 80805 M\"unchen, Germany}
\centerline{\footnotesize E-mail: guenterg@desy.de}
\vspace*{0.3cm}
\centerline{\footnotesize (On behalf of the H1 and ZEUS collaborations)}

\vspace*{0.5cm} 
\abstracts{The expectations for and the measurements
  of transverse energy flows, single particle $p_{T}$ spectra, and
  the rate of forward jets in deep inelastic $ep$~events from the H1
  and ZEUS experiments at HERA are reported and discussed. It is shown
  that together they offer a good chance to establish deviations from
  the DGLAP paradigm.  At the present level of limited statistics the
  measurements are compatible with predictions using BFKL resummation
  and with the color dipole model.  Models based on DGLAP evolution
  describe the $p_{T}$ spectra and forward jets less well but are not
  ruled out yet.}
\normalsize\baselineskip=15pt
\setcounter{footnote}{0}
\renewcommand{\thefootnote}{\alph{footnote}}

\section{Expectations and Observables}

\begin{wrapfigure}{r}{8.5cm}
\vspace{-0.5cm}
\begin{center}
\begin{tabular}{cc}
 \epsfig{file=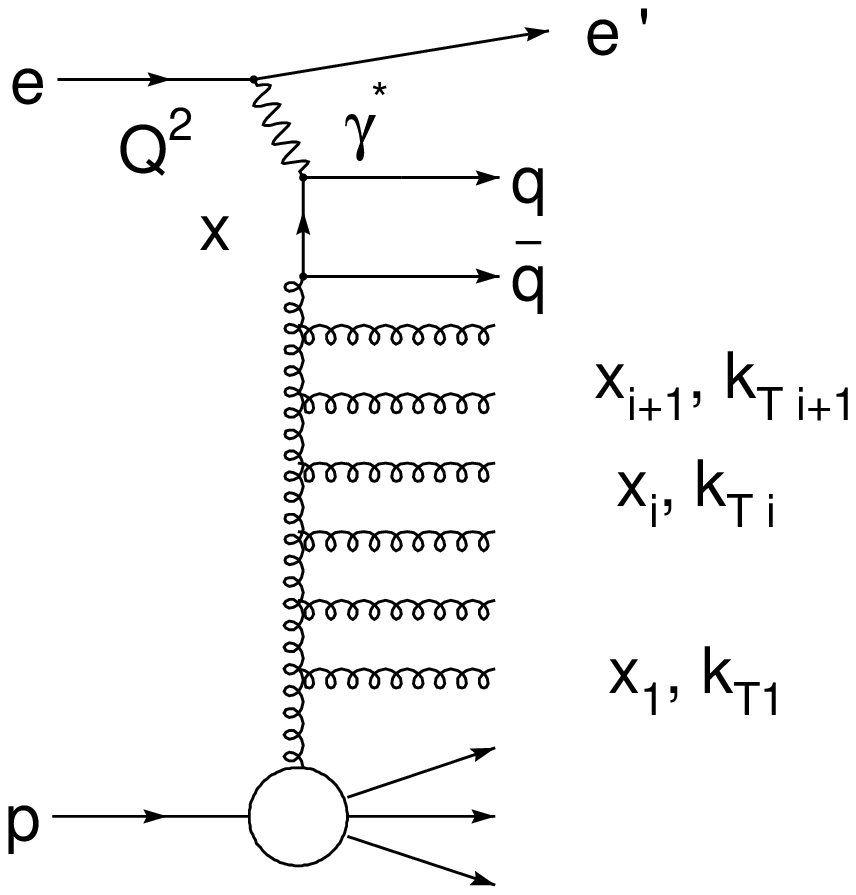,height=5cm,width=6cm}
 \mbox{\hspace{-4cm}
 \epsfig{file=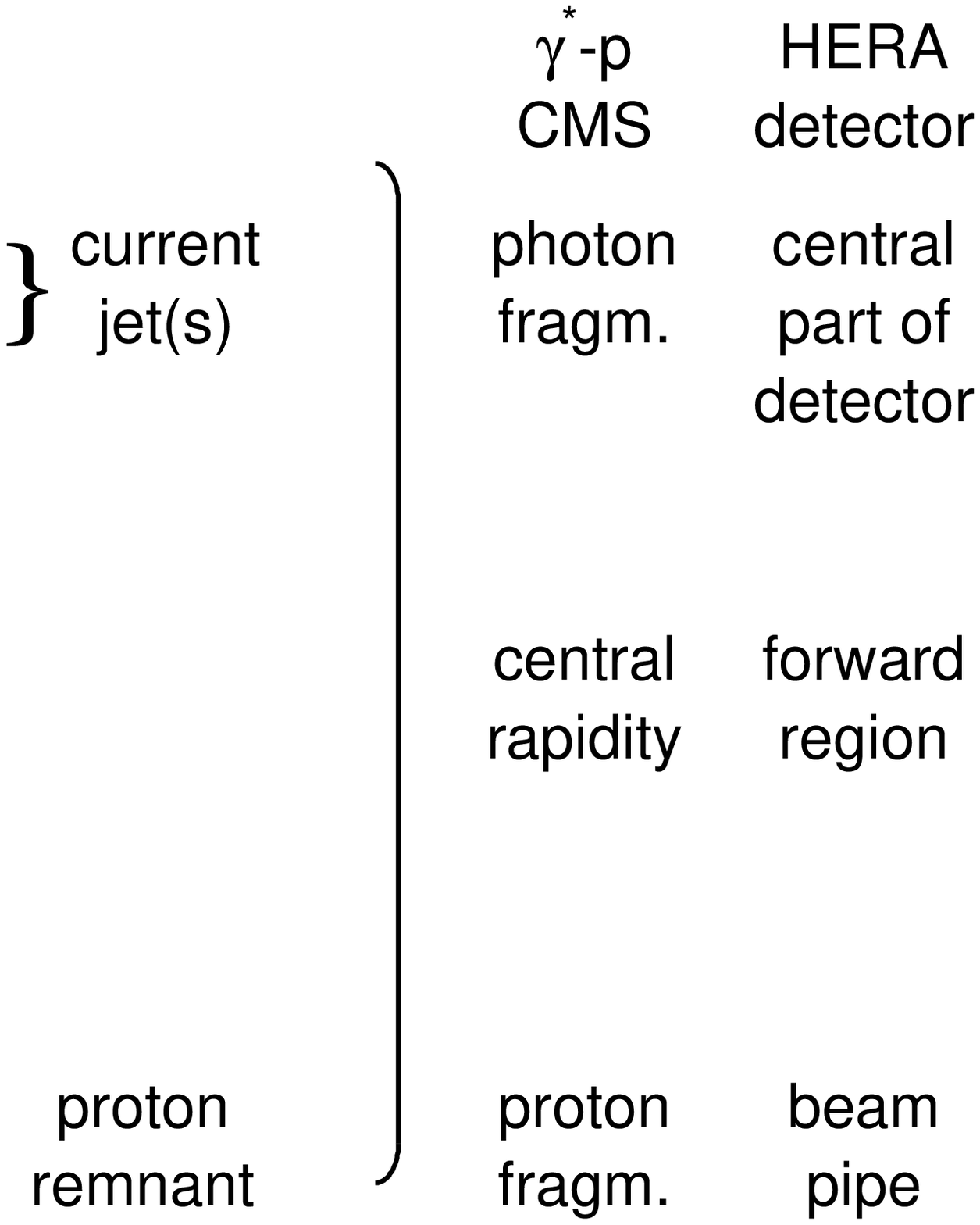,height=5cm,width=5.8cm}
 }
\end{tabular}
\end{center}
\vspace{-0.2cm}
\fcaption{Diagram of an $ep$~collision at low $x$.}
\label{ladder}
\vspace{-0.6cm}
\end{wrapfigure}

In deep inelastic scattering (DIS) at low $x$ the simple picture of
the quark parton model, where a virtual photon interacts with a
point-like parton in the proton and nothing else happens, has to be
modified. The probability that additional partons, particularly
gluons, are radiated increases with decreasing $x$. An example, where
a low $x$ parton, which interacts with the photon, originates from a
parton shower initiated by a gluon with large $x$, is shown in
Fig.~\ref{ladder}. 
In the approximation of just one gluon being
radiated in the initial or the final state, such processes have been
calculated in next to leading order (NLO). Two programs, MEPJET
\cite{mepjet} and DISENT~\cite{disent}, using different methods, are
available, and a new program, DISASTER++~\cite{disaster}, has been
announced at this meeting. These programs generate partons only; no
full event generators exist.

In the leading log approximation any number of more or less collinear
or soft gluons may be radiated. By resumming terms of the form
$(\alpha_s \ln\frac{Q^2}{Q_0^2})^n$ using the DGLAP~\cite{dglap}
evolution equations an initial state parton shower may be evolved from
the scale $Q_0^2$, typically between $0.3$ and $4.0$ GeV$^2$, to the
scale $Q^2$, where the interaction with the photon takes place. DGLAP
evolution implies weakly ordered fractional gluon energies
(longitudinal momenta) $x_i > x_{i+1}$ and strongly ordered gluon
virtualities or transverse gluon momenta $k_{T,i} << k_{T,i+1}$ along
the ladder (Fig.~\ref{ladder}).  Different implementations of
``DGLAP-like'' parton showers and of
matching~\cite{lepto_me_ps,herwig_me_ps} them to leading order (LO)
exact matrix elements for the QCD-Compton (QCDC) and the boson-gluon
fusion (BGF) process exist in LEPTO~\cite{lepto},
RAPGAP~\cite{rapgap}, and HERWIG~\cite{herwig}. For the hadronisation
of the perturbative partons to the observable hadrons, these programs
include phenomenological models.  HERWIG uses the cluster
model~\cite{cluster}; all other programs, including
ARIADNE~\cite{ariadne}, use the LUND string model~\cite{string}, as
implemented in the JETSET~\cite{jetset} code.  These fragmentation
models have been well tested at LEP~\cite{iknowles}. At HERA and at
hadron colliders, the additional complication of proton remnant
fragmentation arises, about which much less is known.

At low $x$, terms of the form $(\alpha_s \ln\frac{1}{x})^n$
become large and can be resummed using the BFKL~\cite{bfkl}
evolution equations. In a physical gauge, these terms correspond to a
ladder diagram in which gluon emissions are strongly ordered in
energies $x_i >> x_{i+1}$, while, in contrast to the DGLAP case, the
transverse gluon momenta perform a random walk along the ladder.
Analytical results at the parton level and first results including
fragmentation functions exist and will be mentioned later.
Unfortunately no implementations of ``BFKL-like'' parton showers are
available in any of the event generators. 

Finally we discuss ARIADNE~\cite{ariadne}, which currently gives the
best overall description of the hadronic final state in DIS, but which
does not fit well into any of the schemes mentioned above. It is based
on the color dipole model (CDM)~\cite{cdm}, where gluon emission
originates from a color dipole stretched between the scattered quark
and the proton remnant.  Further gluon emission leads to a chain of
independently radiating dipoles spanned between color-connected
partons. The first emission is corrected by the QCDC matrix element in
LO\@. 
In ARIADNE, the color charge of the proton remnant is assumed
not to be point-like, leading to a phenomenological suppression of
gluon radiation~\cite{ariadne} in the remnant direction.  The
suppression sets in for hard gluons with wavelengths smaller than the
size of the remnant.  Also the color charge of the scattered quark is
taken to be spread out, depending on the virtuality $Q^2$ of the
photon~\cite{ariadne,cdm_extended_quark}. This in turn leads to a
suppression of radiation in the direction of the scattered quark. An
important feature of ARIADNE is that the probability for a QCDC event
and for any other gluon emission, from a dipole connected to the
remnant at one end, depends on this ``unorthodox
suppression''~\cite{jrathsman}, due to the extended color charge, 
while in ``normal'' QCD, as
implemented in all other programs, it depends on the ratio of parton
densities before and after the emission and on other factors which are 
the same for all programs. At low $x$ this feature leads to a greatly
enhanced rate of QCDC events compared to the other generators. 
Additionally, in 
contrast to the DGLAP-like programs, the gluons in ARIADNE are not
ordered in $k_t$ along the ladder. It has therefore been
argued~\cite{cdm_bfkl} that the partonic final state of ARIADNE is
more closely related to the one expected from BFKL evolution.

It has been shown that all models provide a fair description of basic
event properties~\cite{nbrook_fphera,tcarli_roma,hjung}.

At very low $x$ the gluon density may possibly become so
high in regions (``hot spots'') of the proton that the gluons will no
longer act as free particles but will interact with each other. In
this saturation regime of QCD, the strong coupling is still
small, but the incoherent scattering approximation is no longer valid.
It is not clear whether this new regime could be observed at HERA.

What do we expect to see at HERA\@? Which observables will be most 
sensitive to the underlying parton dynamics?
 
As $x$ decreases below $10^{-3}$ we anticipate deviations from the
DGLAP predictions due to the missing $(\alpha_s \ln\frac{1}{x})^n$
terms.  The most inclusive way to study the structure of the proton is
to measure the probing lepton after the interaction. The resulting
measurement of the proton structure function $F_2$ has the advantages
that it is experimentally relatively easy and that it can be directly
compared to analytical QCD calculations. The rise of $F_2$ with
decreasing $x$ as observed at HERA~\cite{lfavart_rbg} turns out to be
well describable by the DGLAP evolution equations alone, down to $Q^2
\geq 1.0$ GeV$^2$ and $x \geq 10^{-5}$. At this meeting, it was shown
that the $F_2$ data at small $x$ are also well compatible with a
unified BFKL and DGLAP description~\cite{amartin_rbg}.

Perhaps $F_2$ is not sufficiently sensitive to observe the
change in parton dynamics at low $x$. The reasons might be 
that $F_2$ is a too inclusive measurement, averaging out over 
differences only visible in the details of the hadronic final
state. There might not be enough range in $x$ at moderate $Q^2$.
Moreover, the theory has some flexibility due to the freedom of 
choice of the starting parton distributions.    

A greater sensitivity to small $x$ effects might be provided 
by observables based on the hadronic final state emerging from 
the underlying partons. It can be further enhanced by making 
use of the differences between the DGLAP and the BFKL cascade.
However, there is a price to pay. The measurements have to be 
compared to Monte Carlo simulations including the QCD effects, 
the transition from partons to hadrons, the particle decays, 
and detector effects. The observables studied so far are the 
transverse energy flows, single particle $p_T$-spectra, and 
forward jets. 

\section{Transverse Energy Flows}

BFKL dynamics predicts~\cite{jkwiecinski_avet,jbartels_avet}, due to
the lack of ordering of $k_T$ in the parton cascade, more average
transverse energy $E_T$ and harder tails in the $E_T$ distribution for
decreasing $x$ than in the DGLAP scenario.  These observables are
measured calorimetrically in the central rapidity\,\footnote{Rapidity
is here always pseudo-rapidity, i.e. $\eta = -\ln\tan(\theta/2)$.}
\,region, between the struck quark and the proton remnant, in the
hadronic center of mass (hcms), that is the rest system of the
exchanged virtual photon and proton. In that frame all
$E_T$\,\footnote{$E_T=\sum E_i \sin\theta_i$, $i$ runs over
calorimeter cells.} \,is due to either QCD radiation or
hadronisation. As indicated in Fig.~\ref{ladder} the central rapidity
in the hcms corresponds to the forward region
in the frame of the HERA detectors.
\begin{figure}[thb]
 \centering 
 \epsfig{file=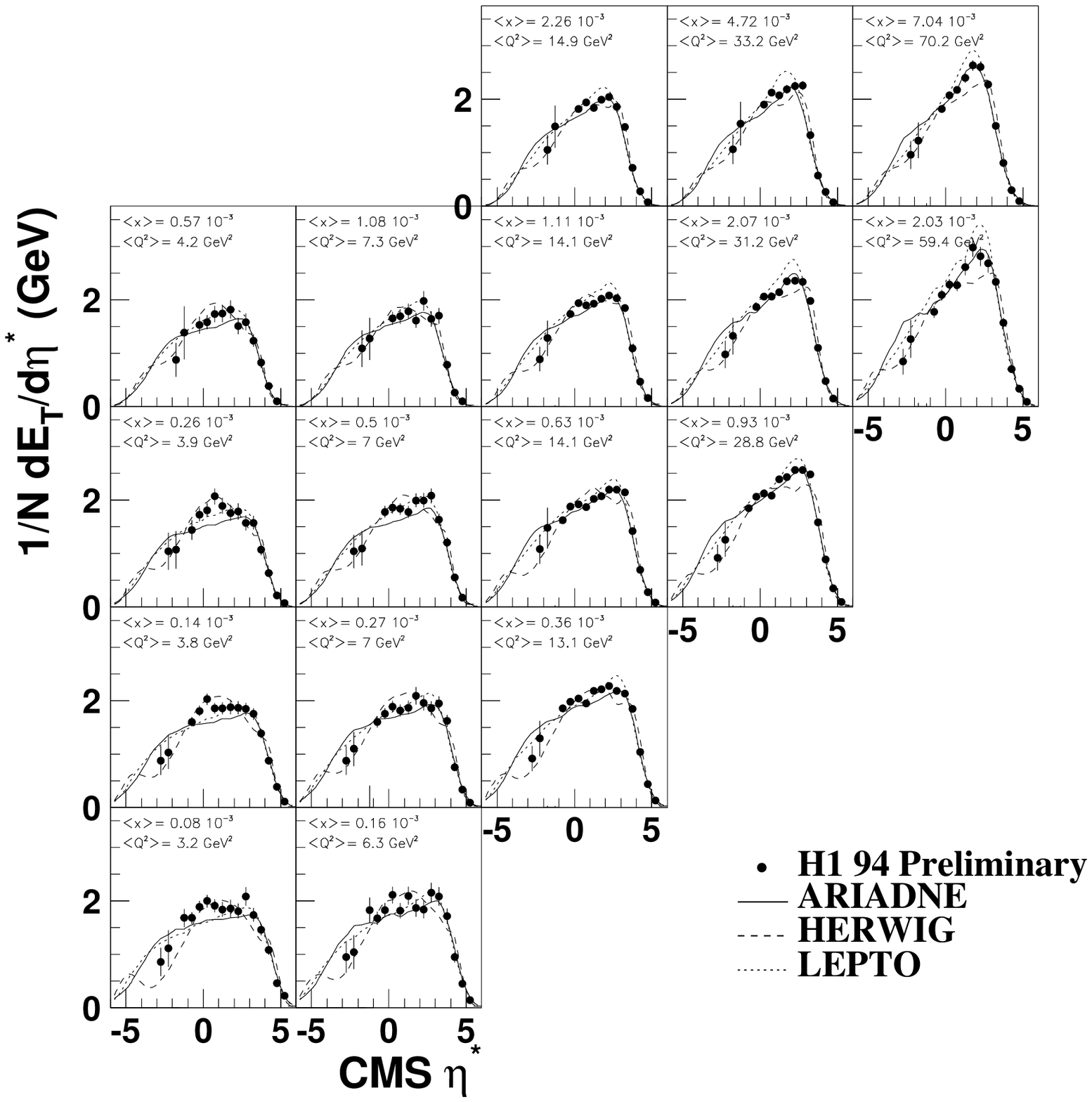,
  height=13cm,width=15cm,clip=}
 \fcaption{Transverse energy flow as a function of rapidity 
  $\eta^*$ in the hcms. The H1 data points show statistical errors
  only, except for two points measured with the plug calorimeter where
  the statistical and systematic errors are given. The lines are the
  Monte Carlo predictions of three different models.} 
 \label{h1_etflow94_prelim}
\end{figure}

First results on the $E_T$-flows have been published by the H1
collaboration~\cite{h1_hfs92,h1_etflow92,h1_etflow93_fwdjets93}. In
ref.~\cite{h1_etflow93_fwdjets93} they have been reported for DIS
events with $5$ GeV$^2 < Q^2 < 50$ GeV$^2$ and with $10^{-4} < x <
10^{-2}$ in the hcms and have been compared to the then current model
predictions of LEPTO 6.1 and ARIADNE 4.03. In the largest $x$ and
$Q^2$ bin, i.e.\ $x \approx 5 \times 10^{-3}$ and $Q^2 \approx 33$
GeV$^2$, the data and the two models were found to agree. With
decreasing $x$ for fixed $Q^2$ the $E_T$ from LEPTO 6.1 was found to
fall more and more below the data for rapidities
away from the current jet, while ARIADNE 4.03
still managed to give a level of $E_T$ in agreement with the data.

At this point in time one could have had the impression that models with
DGLAP-like dynamics like LEPTO fail to describe the data at low $x$,
while models with some BFKL-like features like ARIADNE are
successfull.  However, soon after, two new phenomenological features
have been added to LEPTO which allowed a reasonably good description
of the transverse energy flows also for decreasing $x$.  The new
features are soft color interactions~\cite{gingelman_sci}, with the
intention to describe rapidity gap events without modeling a Pomeron
and its structure function, and a modified sea-quark/remnant
treatment, giving a smoother transition from BGF events to events
where the photon interacts with a sea-quark.

The H1 collaboration in the meantime has produced new preliminary
data~\cite{h1_etflow94,fhess}, covering a larger range in the kinematic
plane from $3 < Q^2 < 70$ GeV$^2$ and $8 \times 10^{-5} < x < 7 \times
10^{-3}$. The extension to lower $Q^2$ was achieved by analysing data
from special runs, where the point of the $ep$~interactions was
shifted from the nominal position in the direction of the proton beam
in order to have access to smaller lepton scattering angles. In
addition, the $E_T$ at two very forward rapidity bins was measured by
H1 using their plug calorimeter ($0.72^{\circ} < \theta_{lab} <
3.3^{\circ}$), which closes the gap between the beam-pipe and the
forward part of their liquid Argon calorimeter.

\begin{figure}[hbt]
 \centering 
 \epsfig{file=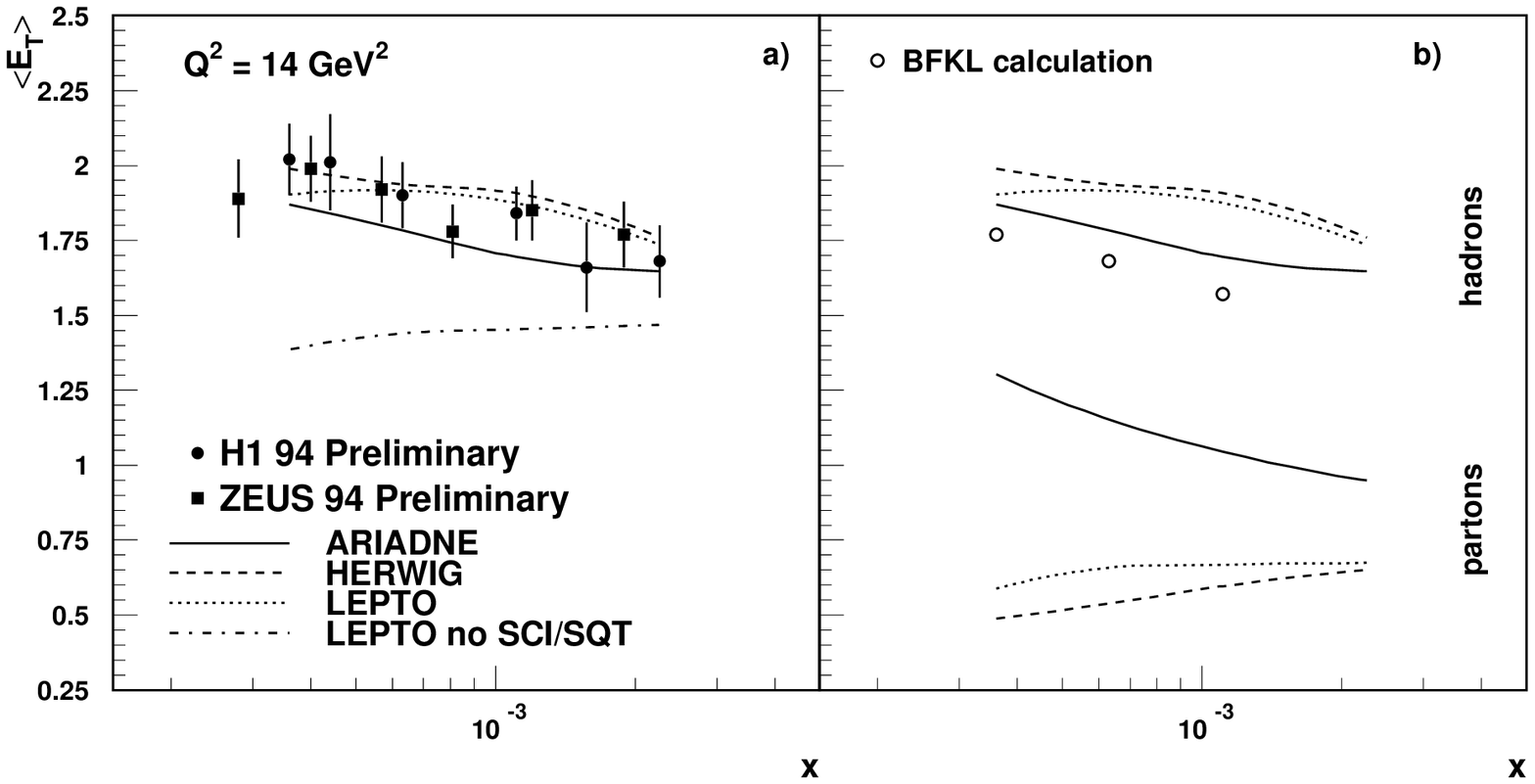,
  height=6cm,width=14cm} 
 \fcaption{Mean transverse energy in the
  central rapidity bin in the hcms for the bin $Q^2 = 14$ GeV$^2$.
  Besides the data from H1 and ZEUS, QCD model predictions (lines)
  are shown for hadrons in a) and for partons in b). An analytic
  BFKL calculation at the parton level is shown as open circles.}
 \label{h1_zeus_kwiecinski_partons_avet94_q214_prelim}
\end{figure}

The data as a function of rapidity\,\footnote{The direction of the
proton is to the left (negative rapidity).} \,in the hcms are shown
in bins of $x$ and $Q^2$ in Fig.~\ref{h1_etflow94_prelim}.
They are compared to predictions from ARIADNE 4.08, HERWIG 5.8, and
LEPTO 6.4. The models give only a fair description of the data over
the large range in $x$ and $Q^2$. The average $E_T$ in the central
rapidity $-0.5 < \eta^* < 0.5$ as a function of $x$ in bins
of fixed $Q^2$ increases with decreasing $x$~\cite{h1_etflow94,fhess}.
This is demonstrated in
Fig.~\ref{h1_zeus_kwiecinski_partons_avet94_q214_prelim}a for the bin
$Q^2 = 14$ GeV$^2$, together with data from ZEUS~\cite{npavel_etflow},
and with predictions 
from different generators at the hadron level.
For LEPTO the predictions are shown with the new soft color
interactions (SCI) and the new sea-quark treatment (SQT) turned on and
off. The data of both experiments are in good agreement and can be
described by all models, DGLAP and BFKL-like.  In
Fig.~\ref{h1_zeus_kwiecinski_partons_avet94_q214_prelim}b the
predictions of the models at the hadron level are contrasted with
those at the parton level and with an analytic parton level BFKL
calculation~\cite{jkwiecinski_avet}. At the parton level the
DGLAP-like models LEPTO and HERWIG show the opposite slope in $x$ than
ARIADNE, the BFKL calculation, the hadron level of all models, and the
data. The BFKL calculation gives the highest transverse energy, $40\%$
to $50\%$ above the partons from the color dipole model.  There are
large differences in the contribution from fragmentation to the total
$E_T$ between the models. It is about $15\%$ for the BFKL calculation
(comparing it to data), $35\%$ for ARIADNE, and about $70\%$ for LEPTO
and HERWIG.
\begin{figure}[htb]
 \centering 
 \epsfig{file=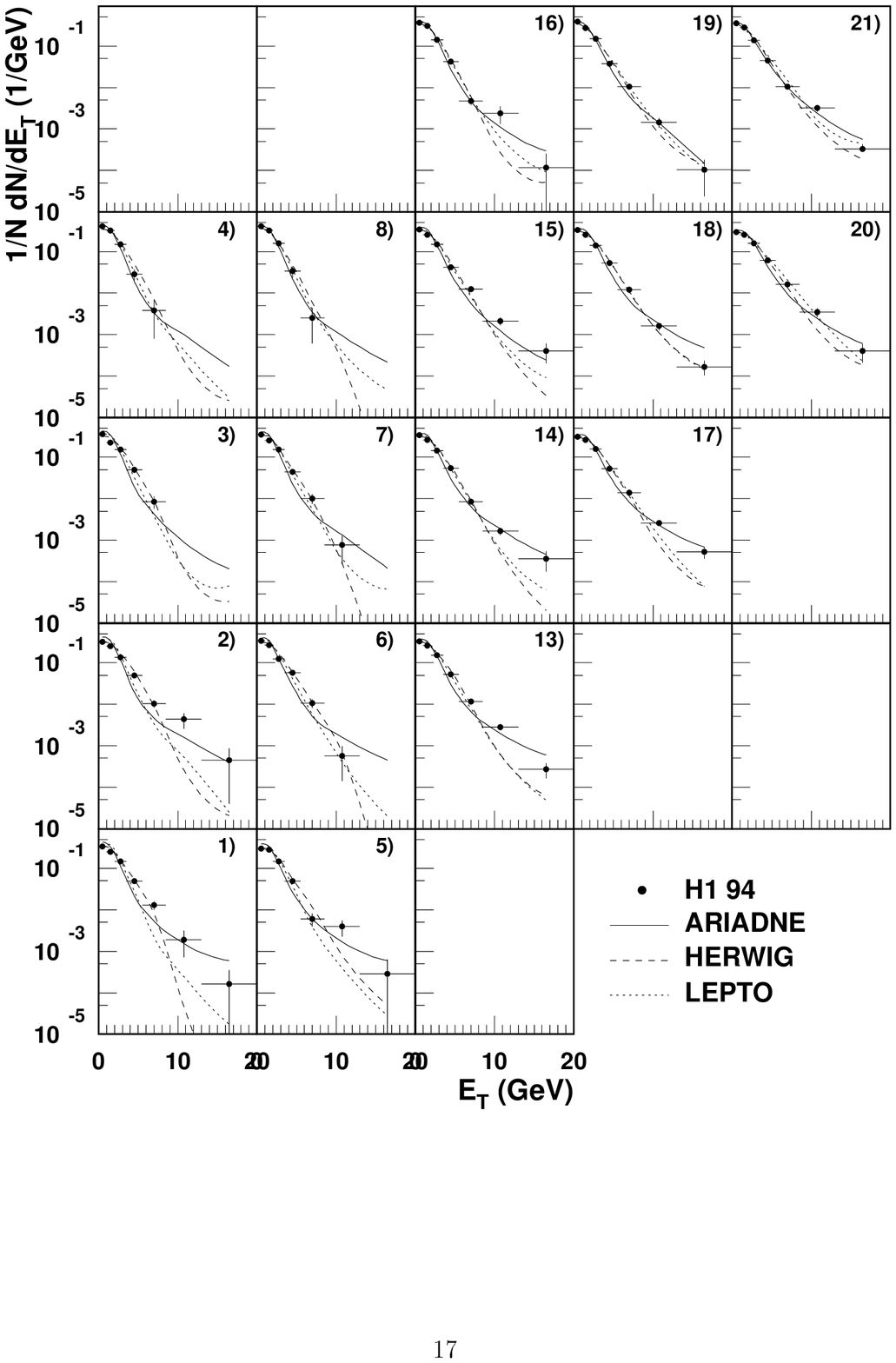,
  height=12cm,width=14cm,
  bbllx=80pt,bblly=180pt,bburx=520pt,bbury=720pt,clip=} 
 \fcaption{Transverse energy distribution
  in the central rapidity bin in the hcms. Only statistical errors
  are shown.}
 \label{h1_models_et94_prelim}
\end{figure}

Concerning the measurement of the mean $E_T$ in the central rapidity
bin, we are now in the situation that all models describe the data.
Although the models differ in their underlying parton dynamics, the
mean $E_T$ can be made to agree by exploiting as yet unconstrained
variations in hadronisation models.

Another observable investigated by H1 is the distribution of the $E_T$
in the central rapidity bin per event. The $E_T$ originating from
hadronisation is expected to be limited, while high values of $E_T$
are more likely to be produced by hard parton radiation.  Preliminary
data~\cite{h1_etflow94,fhess}, in the same bins of $x$ and $Q^2$ as in
Fig.~\ref{h1_etflow94_prelim}, are displayed in
Fig.~\ref{h1_models_et94_prelim} together with predictions from
ARIADNE 4.08, HERWIG 5.8, and LEPTO 6.4.  While at the largest $x$ and
$Q^2$ the data and the models agree, with decreasing $x$ and $Q^2$ the
data and ARIADNE appear to exhibit harder tails than the DGLAP-like
models. As can be seen from the figure a firm conclusion can only be
drawn with more statistics extending the range in $E_T$ to larger
values. Comparing the $E_T$ distribution for the highest and
lowest $x$ value for fixed $Q^2$, one finds that the data have harder
tails~\cite{h1_etflow94,fhess} with decreasing $x$, suggesting a rise
in parton activity.

\section{Single Particle $p_T$ Spectra}

The $E_T$ measured by the calorimeter can be due to many soft
particles, predominantly from hadronisation, or due to a few hard
particles, mainly from hard gluon
\begin{figure}[hbt]
 \centering
 \epsfig{file=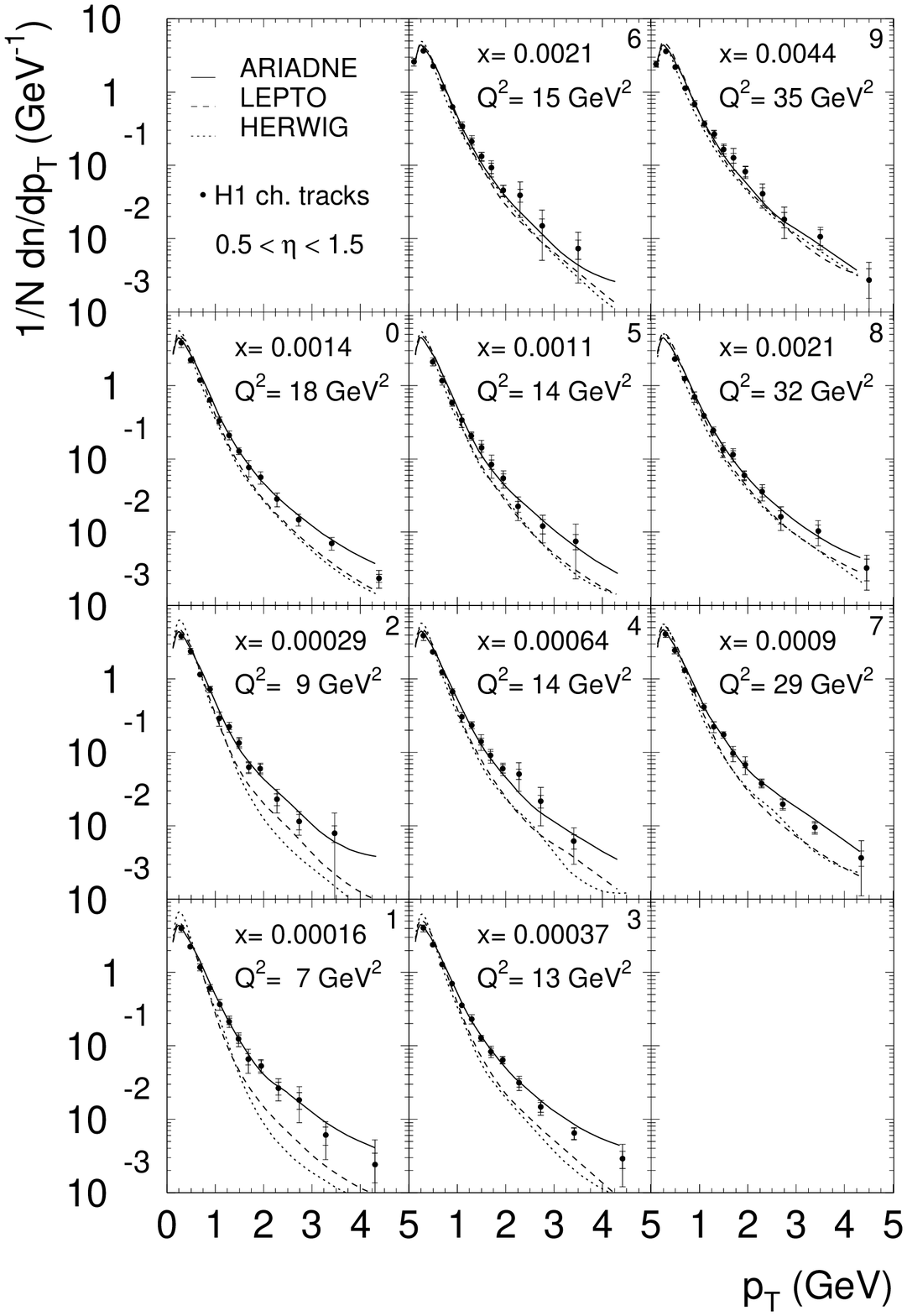,
    height=13cm,width=14cm,
    bbllx=30pt,bblly=20pt,bburx=490pt,bbury=690pt,clip=} 
 \fcaption{The transverse momentum spectra of charged particles in the
    hcms. The data are shown for nine different kinematic bins and 
    the combined sample (bin $0$). Statistical and full errors
    are given.}
 \label{h1_models_pt1}
\end{figure}
emission, or due to both. Therefore it has been
suggested~\cite{mkuhlen_pttail} that the hard tail of the $p_T$
distribution of single charged particles in the hcms might offer
better sensitivity to the basic parton dynamics since it is more
directly linked to hard gluon emissions.

\begin{figure}[htb]
 \centering 
 \epsfig{file=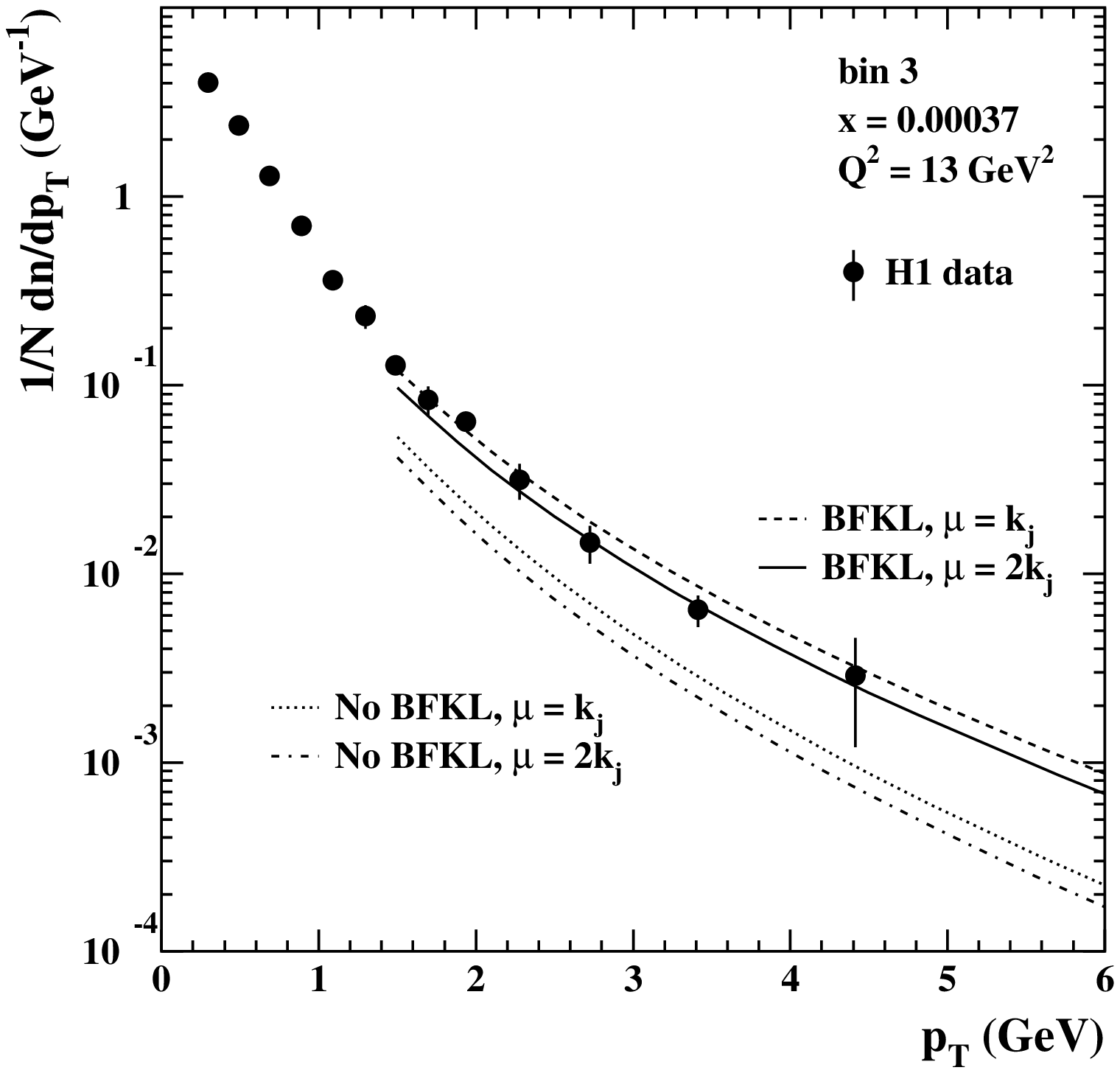,
    height=7.5cm,width=10cm,
    bbllx=50pt,bblly=210pt,bburx=490pt,bbury=645pt,clip=} 
 \fcaption{The transverse momentum spectra of charged particles in the hcms
    for fixed low $x$ and $Q^2$ (bin $3$). The data and predictions
    from a BFKL calculation, with BFKL effects off and on, and with
    two different choices for the scale $\mu$ of the fragmentation
    function are compared.}
 \label{h1_bfkl_pt_bin3}
\end{figure}

The charged particle $p_T$ distributions in bins of $x$ and $Q^2$ as
measured by H1~\cite{h1_pt94} in the rapidity interval from $0.5$ to
$1.5$ are shown in Fig.~\ref{h1_models_pt1}. At large $x$ all three
models presented agree with the data. With decreasing $x$, LEPTO and
HERWIG significantly fall below the data for increasing $p_T$.
Predictions~\cite{jkwiecinski_pt} based on BFKL resummation and
convolution with fragmentation functions for the transition from
partons to charged hadrons~\cite{jbinnewies_fragm} have been made for
the three lowest bins in $x$, i.e.\ bins 1, 2, and 3. As demonstrated
in Fig.~\ref{h1_bfkl_pt_bin3} for bin 3, the data and the BFKL result
agree quite well for $p_T \geq 1.5$ GeV. The absolute normalization
was derived from a comparison of the BFKL calculation of the forward
jet cross section with data from H1 to be discussed later. Also
displayed in the figure is the calculation with BFKL effects turned off,
which falls
below the data. It is apparent from the figure that the
significance of the agreement can be increased with more data at higher
$p_T$. In addition, BFKL effects would become stronger, if the
measurement of charged particles could be extended further in the
direction of the proton remnant, i.e.\ to the rapidity interval $-0.5 <
\eta^* < 0.5$.

\section{Forward Jets}

\begin{wrapfigure}{r}{5cm}
 \begin{center} 
 \epsfig{file=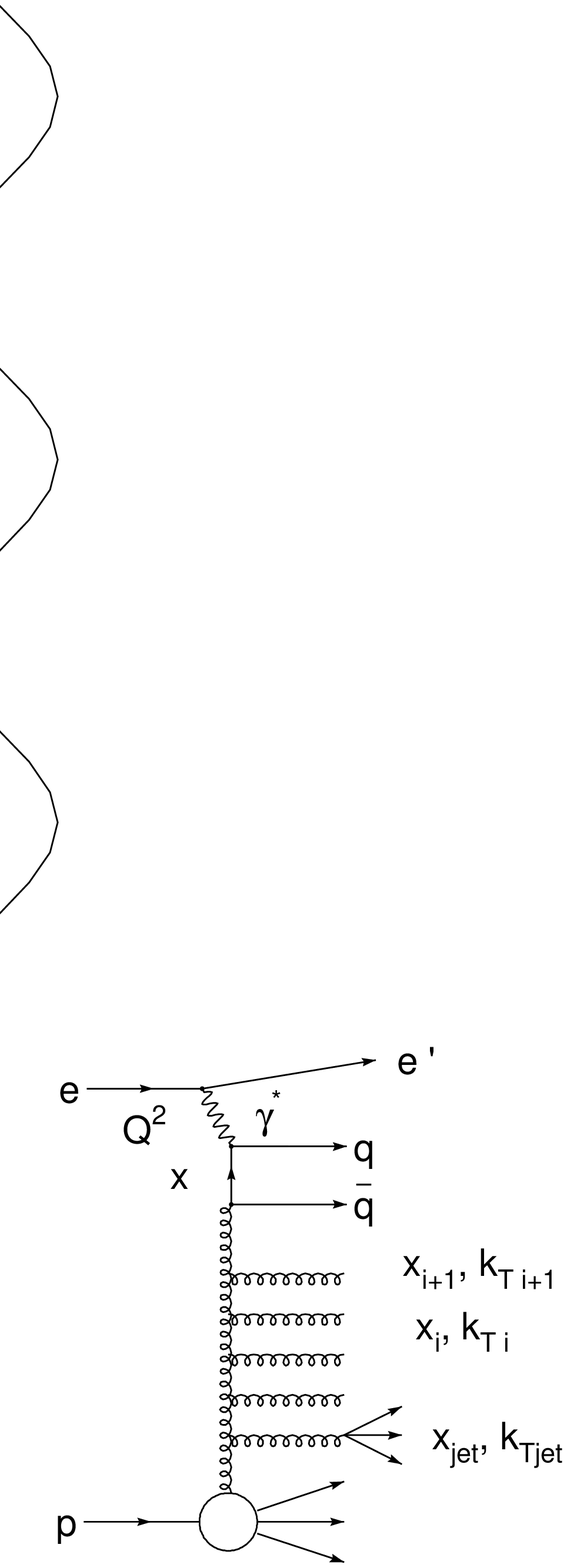,
       height=5cm,width=5cm,clip=}
 \end{center} 
 \fcaption{Parton evolution with forward jet.}
 \label{ladder_fwdjet}
\end{wrapfigure}

The cross section for forward jets in DIS as a function of $x$ has
been advertised~\cite{fwdjets} for some time now as an observable
enhancing the effects of BFKL resummation.  Diagrammatically the
situation is described in Fig.~\ref{ladder_fwdjet}. The forward jet
is defined by the azimuthal angle $\theta_{jet}$ between the jet and
the proton direction and its transverse
momentum $k_{Tjet}$. The energy of the jet is then given by $E_{jet} =
k_{Tjet}/\sin\theta_{jet}$ and $x_{jet} \approx E_{jet}/E_p$. We now
require that $k^2_{Tjet} \approx Q^2$ which suppresses the phase space
for forward jet production and for gluon emission between
the forward and the current jet in the DGLAP case. In addition, we
demand that $x_{jet}/x$ be as large as possible which maximizes the
phase space for the production of a forward jet and gluon radiation
between it and the current jet in the BFKL scenario. For this
particular kinematic configuration the resummation of $\alpha_s \ln
x_{jet}/x$ terms should lead to a sizeable growth of the forward jet
cross section with decreasing $x$.  Another advantage of this
observable is that the parton densities of the proton are probed at
rather large scales, $x_{jet}$ and $k^2_{Tjet} \approx Q^2$, where
they are well known.

Published results on forward jets are available from
H1~\cite{h1_etflow93_fwdjets93} and new preliminary results from
H1~\cite{h1_fwdjets94} and from ZEUS~\cite{zeus_fwdjets94}.  The cuts
for the selection of DIS events and the requirements on the forward
jet are summarized in Table~\ref{selection}. Both experiments use a
cone algorithm, requiring a minimum $p_{Tjet}$ in a cone of radius $R
= \sqrt{\Delta\eta^2 + \Delta\Phi^2}$ in the HERA frame.
\begin{table}[htb]
 \tcaption{DIS event and forward jet selection}
 \label{selection}
 \begin{center}
 \begin{tabular}{||c|c|c||} \hline \hline
               &   H1             &   ZEUS       \\ \hline \hline
 $E_e >$       & $11$ GeV         &   $10$ GeV   \\
 $y_e >$       & $0.1$            &   $0.1$      \\
 $\theta_e <$  & $173^{\circ}$    &   -          \\
 $\theta_e >$  & $160^{\circ}$    &   -          \\ \hline
 cone in lab, R   & $1.0$            &   $1.0$      \\
 $\theta_{jet} >$ & {\boldmath $7^{\circ}$} & 
                    {\boldmath $10^{\circ}$}  \\
 $p_{Tjet} >$     & {\boldmath $3.5$} GeV   & 
                    {\boldmath $5.0$} GeV     \\
 $x_{jet} >$          & $0.035$      & $0.035$ \\
 $\theta_{jet} <$       & $20^{\circ}$      &  -  \\
 $\frac{p^2_{Tjet}}{Q^2} >$ & $0.5$        & $0.5$ \\
 $\frac{p^2_{Tjet}}{Q^2} <$ & {\boldmath $2.0$} & {\boldmath $4.0$} \\
 \hline \hline
 \end{tabular}
 \end{center}
\end{table}
There are differences in the current analyses between the two
experiments in the cuts on $\theta_{jet}$, $p_{Tjet}$, and the upper
limit on $p^2_{Tjet}/Q^2$ which will be discussed later.
With the CDM the transverse energy flow around the forward jet axis
can be well described for different values of
$p_{Tjet}$~\cite{h1_fwdjets94} as well as many other
distributions~\cite{gcontreras,elobodzinska}.

\begin{figure}[htb]
\vspace{-0.5cm}
 \centering 
 \epsfig{file=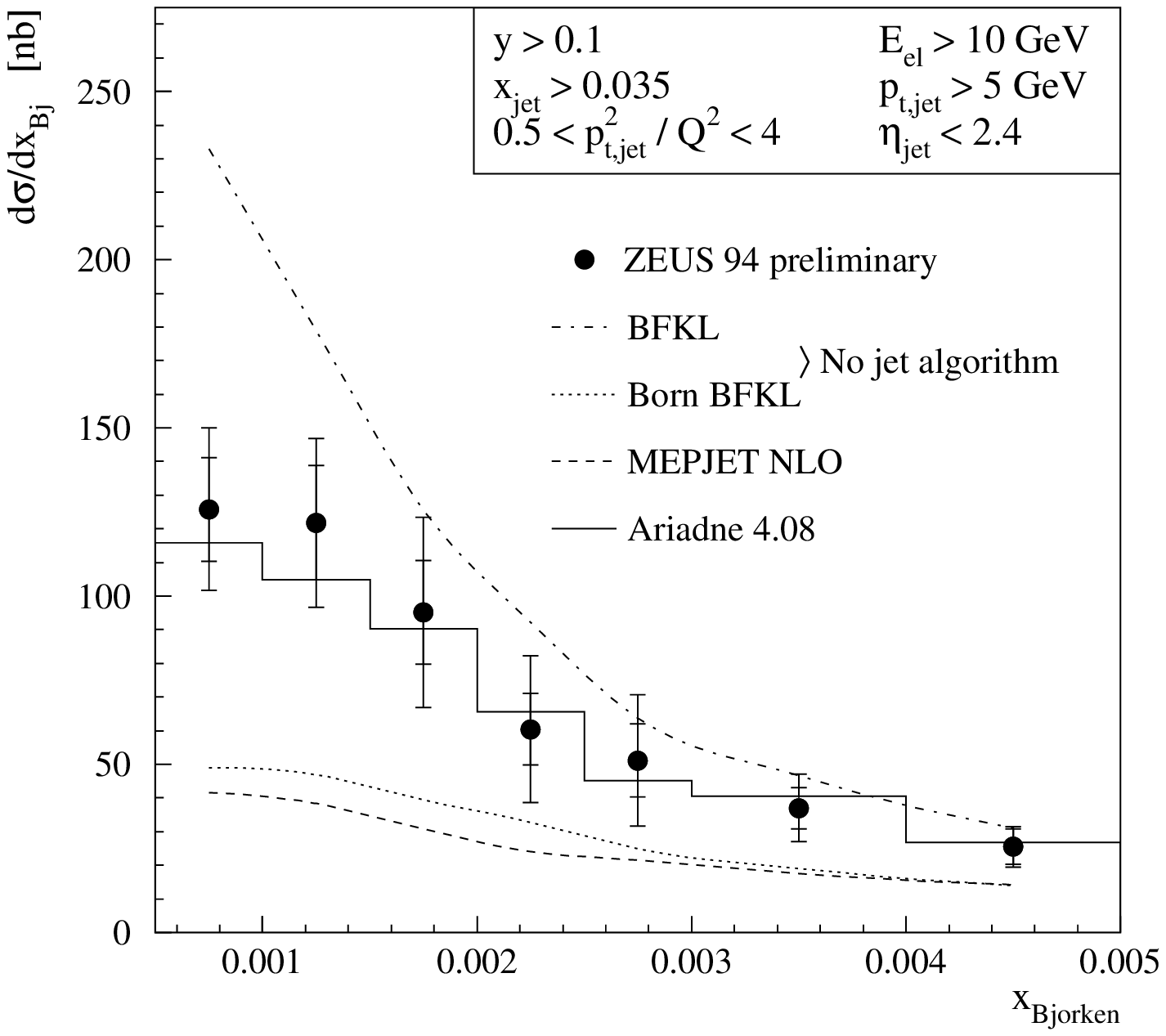,
    height=8cm,width=12cm} 
 \fcaption{The differrential forward jet
    cross section as a function of $x$. The ZEUS data have been
    corrected to the parton level. Statistical and systematic errors
    (not yet complete) are included. BFKL calculations, with BFKL
    effects on (dash-dotted line) and off (dotted line), calculations
    in NLO QCD using MEPJET, and parton level results from the
    color dipole model are shown.}
 \label{zeus_fwdjets2}
\end{figure}

In Fig.~\ref{zeus_fwdjets2} the ZEUS forward jet cross section
corrected to the parton level of ARIADNE is shown and compared to
several parton level calculations: an analytic BFKL
calculation~\cite{jbartels_fwdjets}, the same calculation but without
any gluon emission between the forward and the current jet system
(Born BFKL), and a fixed NLO QCD calculation using MEPJET\@. The data
show a much faster rise with decreasing $x$ than the calculations
without BFKL resummation. The BFKL calculation shows an even more
dramatic rise.  The authors~\cite{jbartels_fwdjets}, however, point
out that several effects which have not been taken into account might
lower their prediction. ZEUS presents their preliminary result
corrected to the parton level in order to compare to parton level
calculations. Using ARIADNE correction factors are found to vary
between $0.6$ for the lowest $x$ bin and $\approx 1$ for larger $x$.
It will be useful to have the ZEUS cross section also at the hadron
level, since it is not clear, what the relationship is between the
parton level of the CDM and BFKL\@. It could also allow comparisons
with H1 data which are corrected to the hadron level.

\begin{figure}[htb]
 \centering 
 \epsfig{file=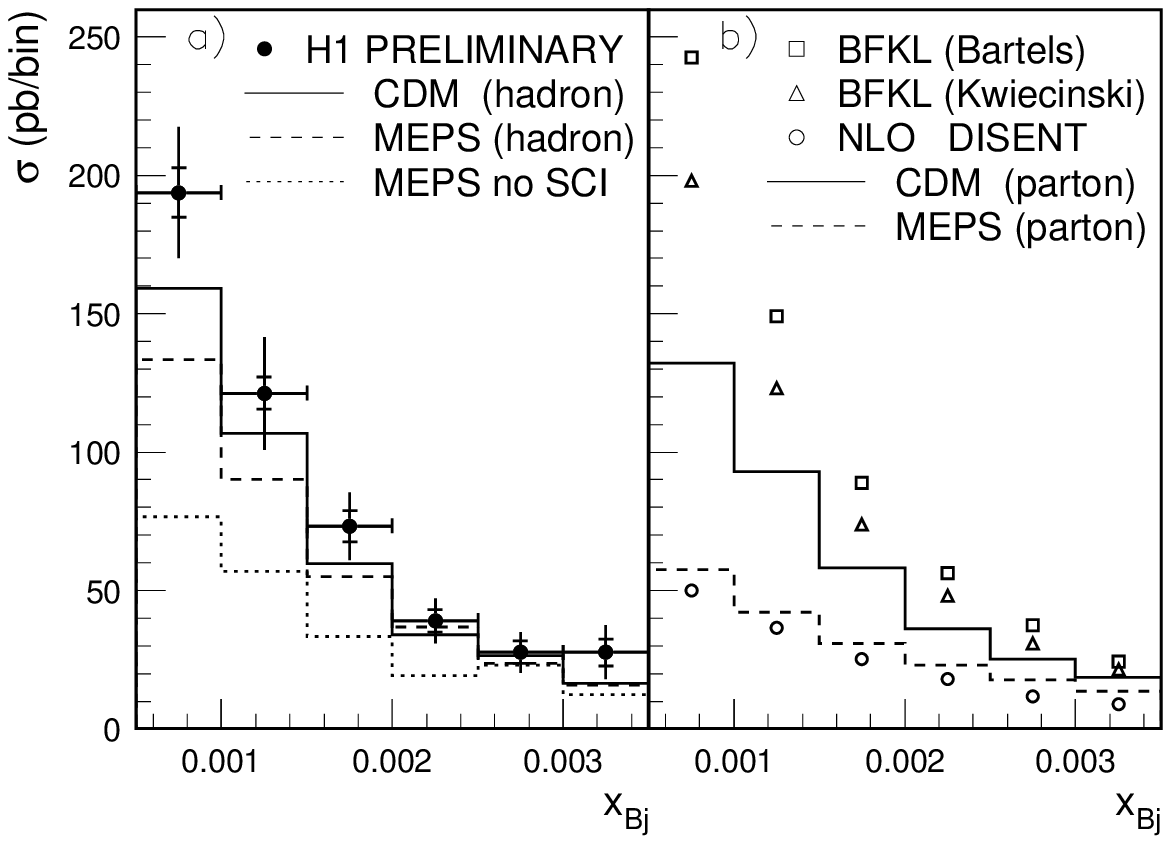,
    height=8cm,width=15cm} 
 \fcaption{The forward jet cross section as
    a function of $x$. The H1 data have been corrected to the hadron
    level. Statistical and systematic errors are included. QCD model
    predictions for hadrons are superimposed in (a) and for partons in
    (b). Two analytic BFKL calculation (open squares and triangles)
    and a NLO calculation using DISENT (open circles) are also shown 
    in (b).}
 \label{h1_fwdjets}
\end{figure}

The preliminary H1 results on the forward cross section are shown in
Fig.~\ref{h1_fwdjets}a and are compared to predictions from LEPTO
(MEPS) with and without soft color interactions (SCI) and to ARIADNE
(CDM). Again a fast rise of the cross section is observed with the CDM
falling only slightly below the data. In Fig.~\ref{h1_fwdjets}b models
and calculations at the parton level are shown: ARIADNE, LEPTO, two
BFKL calculation~\cite{jbartels_fwdjets,jkwiecinski_pt}, and a NLO
calculation using DISENT\@. ARIADNE shows a similar $x$ dependence at
the parton and hadron level with hadronisation effects amounting to
less than $20\%$. The parton level forward jet cross sections of LEPTO
and of DISENT in NLO agree as expected and show a moderate increase.
In LEPTO, with soft color interactions turned on, up to $80\%$ of the
forward jets are created in the hadronisation phase causing LEPTO to
only slightly undershoot the data with decreasing $x$. The BFKL
calculations basically can only predict the dependence on $x$ but not
the normalization. As mentioned before, the computation by Kwiecinski
et al.~\cite{jkwiecinski_pt} fixed the normalization to the data from
H1 while the calculation by Bartels et al.~\cite{jbartels_fwdjets} did
not. Scaling the cross section by Bartels et al.\ down by a factor
$0.8$ brings the two BFKL calculations in agreement.
The color dipole model provides the best description of the data.
However, it should be pointed out that its prediction is rather
sensitive to the power of soft suppression of gluon emission
$(\mu/p_T)^{\alpha}$ due to the proton remnant. The parameter $\alpha$
is related to the dimensionality of the extended proton remnant and is
expected to be $\approx 1$. The H1 forward jets prefer $\alpha \leq
1.0$ while non-jet observables like energy flows prefer $\alpha
\approx 1.5$~\cite{nbrook_fphera}.

One may wonder about the choice for the minimum $p_{Tjet}$, $3.5$ GeV
by H1 and $5.0$ GeV by ZEUS\@. A small value is desirable from the
point of view of statistics and sensitivity to BFKL
effects~\cite{jbartels_fphera}. With increasing $p_{Tjet}$ the slope
in $x$ for the forward jet cross section decreases. The requirement
$p_{Tjet} \approx Q^2$ forces the $Q^2$ to increase, which in turn
increases the minimum $x$ which is probed. On the other hand, to
suppress hadronisation effects, one would want $p_{Tjet}$ to be large.

Another question concerns the choice of the upper limit on
$p^2_{Tjet}/Q^2$ which is $2.0$ for H1 and $4.0$ for ZEUS\@.
Increasing this limit from $2.0$ to $4.0$ increases in the lowest $x$
bins the contribution from $O(\alpha_s)$ matrix elements to the
forward jets by roughly a factor of two. One would also expect an
increased contribution from resolved photoproduction which is not
included in the calculations. After this workshop the author of RAPGAP
has shown~\cite{hjung_fwdjets} that for example the H1 result can also
be described with direct (same as in LEPTO without SCI) and resolved
contributions.

\section{Conclusions}

The tails of the single particle $p_T$ spectra and of the calorimetric
$E_T$ distribution, and the forward jet cross sections offer a good
chance to pin down deviations from the DGLAP paradigm. Other
interesting observables in this quest, like the decorrelation of the
azimuthal angle between the forward jet and the
lepton~\cite{jbartels_fwdjets}, the forward $\pi^{\circ}$ cross
section~\cite{jkwiecinski_fwdpi0}, and the production of more than one
forward jet~\cite{jkwiecinski_2fwdjets} should be pursued.

Along the way, we probably will and have to get a better understanding
of hadronisation effects, particularly of the proton remnant. On the
theory side, higher order effects~\cite{jbartels_rbg} have to be
included in the calculations and a BFKL Monte Carlo
generator~\cite{cschmidt_bfkl_mc,lorr_bfkl_mc} is needed.

In addition to collecting more data at HERA, it would be desirable to
access smaller forward angles for jets~\cite{jbartels_fphera} and
particles and to increase the HERA center of mass energy.
     
\section{Acknowledgements}
 
I want to thank T.\ Carli, M.\ Wobisch, and S.\ W\"olfle for providing
me with plots and J.\ Dainton, J.\ Hartmann, and D.\ Kr\"ucker for a
careful reading of the manuscript. I am also happy to thank the
organizing committee, in particular B.\ Kniel, for the excellent
workshop and the pleasant atmosphere.


\section{References}

\begin{thebibliography}{99}

\bibitem{mepjet}
 E.\ Mirkes and D.\ Zeppenfeld, Phys.\ Lett.\ B380 (1996) 205.

\bibitem{disent}
 S.\ Catani and M.\ Seymour, Proc.\  Workshop on Future Physics at 
 HERA 1995/96, Hamburg, eds.\ G.\ Ingelman, A.\ De Roeck, and R.\ Klanner,
 vol.\ 1 (1996) 519, CERN-TH/96-240 (1996) [hep-ph/9609521], and
 Acta Phys.\ Polon.\ B28 (1997) 863.
 
\bibitem{disaster}
 D.\ Graudenz, these proceedings.
 
\bibitem {dglap}
 V.\,N.\ Gribov and L.\,N.\ Lipatov, Sov.\ J.\ Nucl.\ Phys.\ 15 (1972)
 438 and 675;\\
 G.\ Altarelli and G.\ Parisi, Nucl.\ Phys.\ B126 (1977) 298;\\
 Yu.\,L.\ Dokshitzer, Sov.\ Phys.\ JETP 46 (1977) 641.

\bibitem{lepto_me_ps}
 M.\ Bengtsson, G.\ Ingelman, and T.\ Sj\"ostrand, Proc.\ HERA 
 Workshop 1987, Hamburg, ed.\ R.\,D.\ Peccei, vol.\ 1 (1988) 149;\\
 M.\ Bengtsson and T.\ Sj\"ostrand, Z.\ Phys.\ C37 (1988) 465.

\bibitem{herwig_me_ps}
 M.\, H.\ Seymour, LU-TP-94-12, Nucl.\ Phys.\ B436 (1995) 443 and 
 Comput.\ Phys.\ Commun.\ 90 (1995) 95.
   
\bibitem{lepto} 
 G.\ Ingelman, Proc.\ Workshop on Physics at HERA 1991, Hamburg,
 eds.\ W.\ Buchm\"uller and G.\ Ingelman, vol.\ 3 (1992) 1366;\\
 G.\ Ingelman, A.\ Edin, and J.\ Rathsman, Comput.\ Phys.\ Commun.\ 101
 (1997) 108.

\bibitem {rapgap}
 H.\ Jung, Comput.\ Phys.\ Commun.\ 86 (1995) 147.
  
\bibitem{herwig}
 G.\ Marchesini et al., Comput.\ Phys.\ Commun.\ 67 (1992) 465. 

\bibitem{cluster}
 B.\,R.\ Webber, Nucl.\ Phys.\ B238 (1984) 492;\\
 G.\ Marchesini and B.\,R.\ Webber, Nucl.\ Phys.\ B310 (1988) 461. 

\bibitem{ariadne}
 L.\ L\"onnblad,
 Comput.\ Phys.\ Comm.\ 71 (1992) 15 and Z.\ Phys.\ C65 (1995) 285.

\bibitem{string}
 B.\ Andersson, G.\ Gustafson, G.\ Ingelman, and T.\ Sj\"ostrand, 
 Phys.\ Rep.\ 97 (1983) 31.

\bibitem{jetset}
 T.\ Sj\"ostrand, Comput.\ Phys.\ Commun.\ 82 (1994) 74;\\
 T.\ Sj\"ostrand, PYTHIA 5.7 and JETSET 7.4, CERN-TH 7112/93 (1993)
 and LU TP 95-20 (1995).

\bibitem{iknowles}
 I.\,G.\ Knowles et al., Proc.\ LEP2 Physics Workshop 1995 [hep-ph/9601212].

\bibitem{bfkl}
 E.\,A.\ Kuraev, L.\,N.\ Lipatov, and V.\,S.\ Fadin, Sov.\ Phys.\ JETP
 45 (1977) 199;\\
 Ya.\,Ya.\ Balitsky and L.\,N.\ Lipatov, Sov.\ J.\ Nucl.\ Phys.\ 28 (1978) 822.

\bibitem{cdm}
 G.\ Gustafson, Phys.\ Lett.\ B175 (1986) 453; \\
 G.\ Gustafson and U.\ Petterson, Nucl.\ Phys.\ B306 (1988); \\
 B.\ Andersson, G.\ Gustafson, L.\ L\"onnblad, and U.\ Petterson,
 Z.\ Phys.\ C43 (1989) 625.

\bibitem{cdm_extended_quark}
 L.\ L\"onnblad, M.\ Seymour, et al.,
 Proc.\ Physics at LEP 2 Workshop, eds. G.\  Altarelli,
 T.\ Sj\"ostrand, and F.\ Zwirner, CERN/96-01, vol.\ 2, (1996) 187.

\bibitem{jrathsman}
 J.\ Rathsman, Phys.\ Lett.\ B393 (1997) 181.

\bibitem{cdm_bfkl}
 A.\,H.\ Mueller, Nucl.\ Phys.\ B415 (1994) 373.

\bibitem{nbrook_fphera}
 N.\ Brook et al., Proc.\  Workshop on Future Physics at 
 HERA 1995/96, Hamburg, eds.\ G.\ Ingelman, A.\ De Roeck, and R.\ Klanner,
 vol.\ 1 (1996) 613.

\bibitem{tcarli_roma}
 T.\ Carli, Proc.\ 4th Int. Workshop on Deep Inelastic 
 Scattering and Related Phenomena (DIS 96), Roma, eds. G.\ D'Agostini 
 and A.\ Nigro, (1997) 415.

\bibitem{hjung}
 H.\ Jung, private communication.
%
%

\bibitem{lfavart_rbg}
 L.\ Favart, these proceedings.

\bibitem{amartin_rbg}
 A.\,D.\ Martin, these proceedings.

\bibitem{jkwiecinski_avet}
 K.\ Golec-Biernat, J.\ Kwiecinski, A.\,D.\ Martin, and P.\,J.\ Sutton,
 Phys.\ Rev.\ D50 (1994) 217 and Phys.\ Lett.\ B335 (1994) 220.

\bibitem{jbartels_avet}
 J.\ Bartels, H.\ Lotter, and M.\ Vogt, Phys.\ Lett.\ B373 (1996) 215. 

\bibitem{h1_hfs92}
 H1 Collaboration, T.\ Ahmed et al., Phys.\ Lett.\ B298 (1993) 469.

\bibitem{h1_etflow92}
 H1 Collaboration, I.\ Abt et al., Z.\ Phys.\ C63 (1994) 377.   

\bibitem{h1_etflow93_fwdjets93}
 H1 Collaboration, S.\ Aid et al., Phys.\ Lett.\ B356 (1995) 118.

\bibitem{gingelman_sci}
 A.\ Edin, G.\ Ingelman, and J.\ Rathsman, Z.\ Phys.\ C75 (1997) 57 and
 Phys. Lett. B366 (1996) 371.

\bibitem{h1_etflow94}
 H1 Collaboration, C.\ Adloff et al., contrib. paper pa02-073,
 ICHEP '96, Warsaw, Poland, July 1996.

\bibitem{fhess}
 M.\,F.\ Hess, Ph.D.\ Thesis, University of Hamburg (1996).

\bibitem{npavel_etflow}
 N.\,A.\ Pavel, Proc.\ 4th Int. Workshop on Deep Inelastic 
 Scattering and Related Phenomena (DIS96), Roma, eds. G.\ D'Agostini 
 and A.\ Nigro, (1997) 502.

\bibitem{mkuhlen_pttail}
 M.\ Kuhlen, Phys.\ Lett.\ B382 (1996) 441.

\bibitem{h1_pt94}
 H1 Collaboration, C.\ Adloff et al., Nucl.\ Phys.\ B485 (1997) 3.

\bibitem{jkwiecinski_pt}
 J.\ Kwiecinski, S.\,C.\ Lang, and A.\,D.\ Martin, DTP-97-56 (1997)
 [hep-ph/9707240].

\bibitem{jbinnewies_fragm}
 J.\ Binnewies, B.\,A.\ Kniehl, and G.\ Kramer, Phys.\ Rev.\ D52 (1995)
 4947.

\bibitem{fwdjets}
 A.\,H.\ Mueller, Nucl.\ Phys.\ B (Proc.\ Suppl.) 18C (1990) 125 and
 J.\ Phys.\ G17 (1991) 1443;\\
 W.\,K.\ Tang, Phys.\ Lett.\ B278 (1992) 363;\\
 J.\ Bartels, A.\ De Roeck, and M.\ Loewe, Z.\ Phys.\ C54 (1992) 635;\\
 A.\ De Roeck, Nucl.\ Phys.\ B (Proc.\ Suppl.) 29A (1992) 61;\\
 J.\ Kwiecinski, A.\,D.\ Martin, and P.\,J.\ Sutton, Phys.\ Lett.\ B287
 (1992) 254, Nucl.\ Phys.\ B (Proc.\ Suppl.) 29A (1992) 67, and
 Phys.\ Rev.\ D46 (1992) 921.

\bibitem{h1_fwdjets94}
 J.\,G.\ Contreras, Proc.\ XXXI Rencontres de Moriond in QCD and
 High Energy Hadronic Interactions, ed.\ J.\ Tran Thanh Van, Edition
 Frontieres (1996) 411;\\
 H1 Collaboration, C.\ Adloff et al., contrib. paper pa03-049,
 ICHEP '96, Warsaw, Poland, July 1996.

\bibitem{zeus_fwdjets94}
 S.\ W\"olfle, to be published in Proc.\ 5th Int.\ Workshop 
 on Deep Inelastic Scattering and QCD, Chicago, April 1997, 
 eds. D.\ Krakauer and J.\ Respond. 

\bibitem{gcontreras}
 J.\,G.\ Contreras, Ph.D.\ Thesis, University of Dortmund (1997).

\bibitem{elobodzinska}
 E.\,M.\ Lobodzinska, Ph.D.\ Thesis, Inst.\ of Nucl.\ Physics, Cracow (1997). 

\bibitem{jbartels_fwdjets}
 J.\ Bartels et al., Phys.\ Lett.\ B384 (1996) 300. 


\bibitem{jbartels_fphera}
 J.\ Bartels, A.\ De Roeck, and M.\ W\"usthoff, Proc.\  Workshop on 
 Future Physics at HERA 1995/96, Hamburg, eds.\ G.\ Ingelman, 
 A.\ De Roeck, and R.\ Klanner, vol.\ 1 (1996) 598.

\bibitem{hjung_fwdjets}
 H.\ Jung, to be published in Proc.\  Madrid Workshop on Low $x$ 
 Physics, Miraflores de la Sierra, June 1997.

\bibitem{jkwiecinski_fwdpi0}
 J.\ Kwiecinski, S.\,C.\ Lang, and A.\,D.\ Martin, Phys.\ Rev.\ D55 (1997) 1273.

\bibitem{jkwiecinski_2fwdjets}
 J.\ Kwiecinski, C.\,A.\,M.\ Lewis, and A.\,D.\ Martin, DTP-97-58 (1997)
 [hep-ph/9707375].

\bibitem{jbartels_rbg}
 J.\ Bartels, these proceedings.

\bibitem{cschmidt_bfkl_mc}
 C.\,R.\ Schmidt, Phys.\ Rev.\ Lett.\ 78 (1997) 4531.

\bibitem{lorr_bfkl_mc}
 L.\,H.\ Orr and W.\,J.\ Stirling, DTP-97-48 (1997) [hep-ph/9706529],
 to be published in Phys.\ Rev.\ D\@.

\end{thebibliography}
\end{document}